\documentclass[journal,twoside]{IEEEtran}
\usepackage{amsmath,amsfonts}
\usepackage{algorithm}
\usepackage{algpseudocode}
\usepackage{array}
\usepackage[caption=false,font=normalsize,labelfont=sf,textfont=sf]{subfig}
\usepackage{textcomp}
\usepackage{stfloats}
\usepackage{url}
\usepackage{verbatim}
\usepackage{graphicx}
\usepackage{cite}
\usepackage{float}
\usepackage{graphicx}
\usepackage{subfig}
\usepackage{enumitem}
\usepackage{amsmath} 
\usepackage{cleveref}
\usepackage{caption}
\usepackage{url}
\usepackage{multirow}
\usepackage{booktabs}
\usepackage[table]{xcolor} 
\definecolor{lightgray}{gray}{0.9}

\begin{document}

\title{SparKV: Overhead-Aware KV Cache Loading for Efficient On-Device LLM Inference}

\author{Hongyao Liu,
        Liuqun Zhai,
        Junyi Wang,
        Zhengru Fang,
        Jingshu Chen,
        and Jun Huang
\thanks{H. Liu (hongyaliu4-c@my.cityu.edu.hk), L. Zhai, J. Wang, Z. Fang, and J. Huang (jun.huang@cityu.edu.hk) are with the Department of Computer Science, City University of Hong Kong, Hong Kong SAR. J. Chen is with the Department of Computer Science, University of Alabama in Huntsville. Jun Huang is the corresponding author.}%
}

\maketitle

\begin{abstract}
Efficient inference for on-device Large Language Models (LLMs) remains challenging due to limited hardware resources and the high cost of the prefill stage, which processes the full input context to construct Key-Value (KV) caches. We present SparKV, an adaptive KV loading framework that combines cloud-based KV streaming with on-device computation. SparKV models the cost of individual KV chunks and decides whether each chunk should be  streamed or computed locally, while overlapping the two execution paths to reduce latency. To handle fluctuations in wireless connectivity and edge resource availability, SparKV further refines offline-generated schedules at runtime to rebalance communication and computation costs. Experiments across diverse datasets, LLMs, and edge devices show that SparKV reduces Time-to-First-Token by 1.3$\times$-5.1$\times$ with negligible impact on response quality, while lowering per-request energy consumption by 1.5$\times$ to 3.3$\times$, demonstrating its robustness and practicality for real-world on-device deployment.
\end{abstract}

\begin{IEEEkeywords}
On-device LLM, KV cache loading, overhead-aware, time-to-first-token
\end{IEEEkeywords}

\maketitle

\section{Introduction}
\label{sec:intro}

Large Language Models (LLMs) have achieved strong success in a wide range of commercial applications~\cite{llama3herd2024,openai2023gpt4,geminiteam2023gemini}. At the same time, open-source models such as Qwen~\cite{qwen3technicalreport} and Mistral~\cite{jiang2023mistral} are making LLM inference increasingly practical on edge platforms. This trend is particularly important for context-intensive tasks that rely on private user data, motivating local inference on mobile and edge devices~\cite{asplos25fast, ISCA25aif, TMC24edgellm, IoTJ2025edge,cai2024edge,liu2021switchable,liu2020research}. While the Internet of Things (IoT) spans a wide spectrum of hardware, this work focuses on edge AI computing platforms rather than ultra-low-power microcontrollers. Representative deployment targets include edge AI gateways, in-vehicle computing systems, and smartphones, which can run quantized on-device LLMs but remain resource-constrained when performing LLM inference with intensive context reuse.

Appending reused context to user prompts enhances response accuracy \cite{liu2024cachegen}, making it a critical strategy for applications where key information is repeatedly referenced. Examples include private document analysis, multi-turn conversations, UI navigation histories, and personal cloud media. However, efficiently supporting such context reuse remains a major bottleneck during the on-device LLM prefill stage, where the model processes the entire input context to construct the key-value (KV) cache used for subsequent decoding. On edge devices, the prefill stage imposes heavy demands on computation and memory bandwidth and is widely recognized as the  dominant contributor to Time-to-First-Token (TTFT). We therefore use TTFT~\cite{liu2024cachegen, sosp2023vllm, dao2023flashattention} as the primary metric. For example, a mobile AI agent driven by openClaw~\cite{openclaw} may need to retrieve and analyze a high-definition video from a user's cloud gallery while preserving data privacy through local processing. The resulting multimodal context can easily exceed 15K tokens and may reach 40K tokens \cite{qwen3technicalreport}. On our Redmi K80 Pro with 16\,GB memory, sequential prefill for such workloads takes over 15\,s, which severely degrades interactivity. More broadly, as intelligent agents become increasingly integrated into mobile and IoT ecosystems, many applications involve large amounts of private contextual data that must be processed locally under zero-trust privacy requirements. Then, reducing this latency is essential for practical deployment.

\begin{figure}
    \centering
    \includegraphics[width=0.9\linewidth]{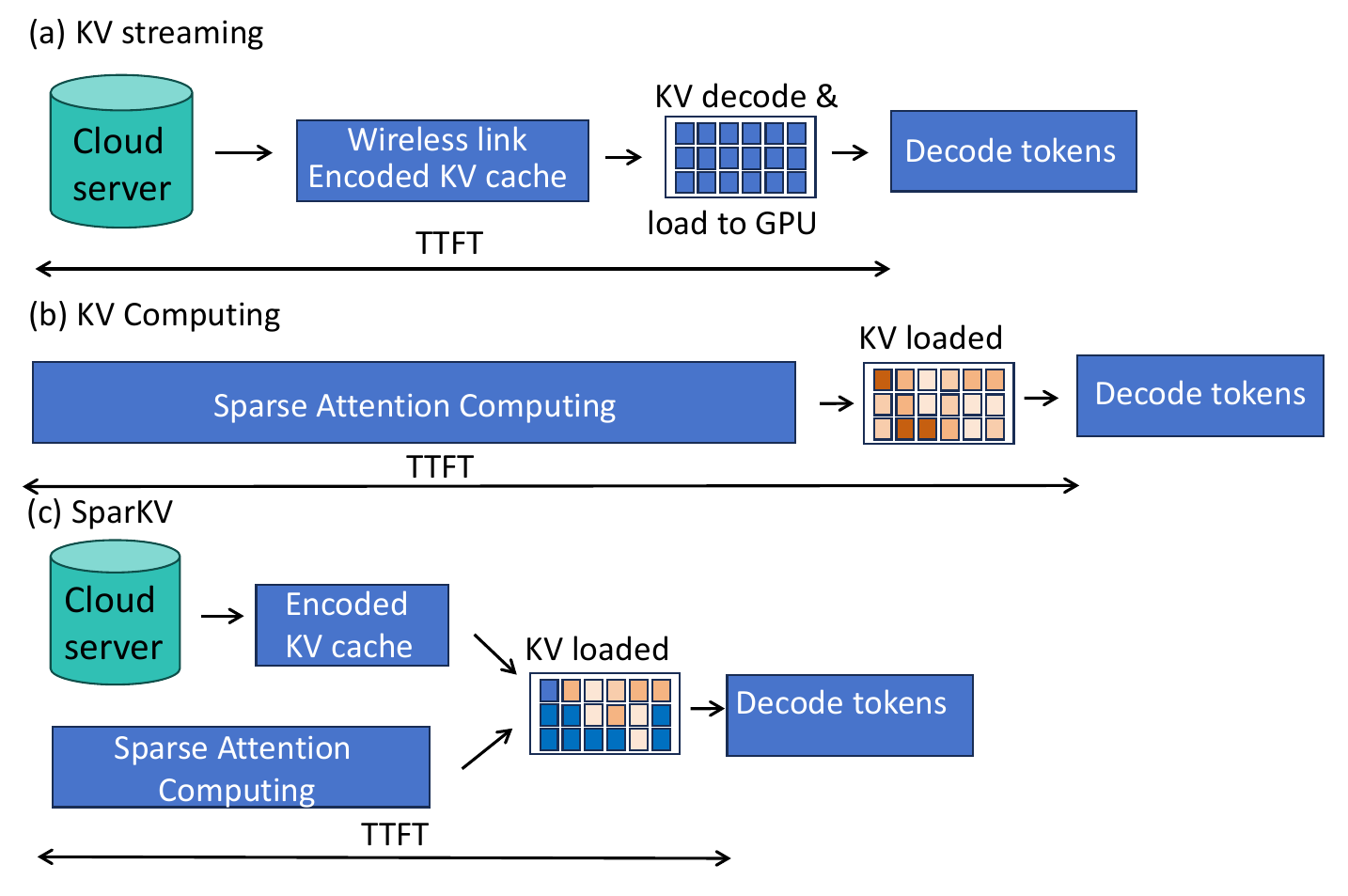}
    \vspace{-0.1in}
    \caption{\small KV cache loading strategies for TTFT reduction: (a) streaming only; (b) computation only; (c) overhead-aware hybrid loading.}
    \vspace{-0.3in}
    \label{fig:kv_loading}
\end{figure}

Existing efforts to reduce TTFT mainly follow two directions: streaming compressed KV cache and accelerating local prefill computation. Fig.~\ref{fig:kv_loading} summarizes three representative workflows. As shown in Fig.~\ref{fig:kv_loading}(a), streaming-based approaches target the I/O bottleneck through KV cache compression. KIVI~\cite{liu2024kivi} quantizes keys and values differently, while H2O~\cite{zhang2023h2o} and LLMLingua~\cite{jiang2023llmlingua} remove redundant tokens from prompts and KV caches. However, the memory footprint remains substantial: even after compression, a 100K-token KV cache for Qwen3-8B~\cite{qwen3technicalreport} still requires 2.5--4\,GB. To mitigate this storage overhead, recent systems store KV caches in the cloud and stream them to edge devices on demand. InfiniGen~\cite{lee2024infinigen} and IMPRESS~\cite{chen2025impress} prioritize KV segments on a per-layer basis, while CacheGen~\cite{liu2024cachegen} adapts compression using layer-wise quantization sensitivity. As shown in Fig.~\ref{fig:kv_loading}(b), a second line of work targets the computation bottleneck of prefill stage. These approaches accelerate local processing through sparse attention~\cite{zhangspargeattention,nips24minference}, hardware-aware kernels~\cite{dao2023flashattention}, and compact attention designs such as Grouped-Query Attention (GQA)~\cite{ainslie2023gqa}. Nevertheless, even with sparse attention~\cite{zhangspargeattention}, our Jetson Orin still requires more than 5\,s to process a 15K-token context. Together, these observations suggest that minimizing TTFT requires jointly exploiting both streaming and computation opportunities.

A recent study~\cite{computeorload} reduces prefill latency through a ``bidirectional convergence'' strategy that computes early context chunks while loading later ones in parallel. However, this approach has three fundamental limitations. First, \textit{rigid computation dependencies}: Transformer computation follows strict layer-wise and causal dependencies over historical tokens, and coarse KV partitioning fails to expose fine-grained opportunities for interleaving computation and streaming. Second, \textit{workload agnosticism}:  KV chunks can incur substantially different streaming and computation overheads, yet existing schemes ignore this heterogeneity, leading to suboptimal schedules for context reuse. Third, \textit{sensitivity to edge volatility}: methods designed for stable server interconnects such as PCIe do not adapt well to the high variability of wireless I/O and fluctuating edge compute capacity, making synchronization between the two paths fragile.

Our key insight is that context chunks should not be overlapped blindly; instead, they should be routed according to their processing overheads and dependency structure. As illustrated in Fig.~\ref{fig:kv_loading}(c), chunks that are cheaper to stream are assigned to the cloud path, while those that are cheaper to compute are processed locally, with the two paths overlapped whenever dependencies permit. Realizing this design requires addressing two challenges: respecting strict token-wise and layer-wise dependencies, and adapting to volatile wireless throughput and dynamic edge resource availability.

In this paper, we present \textit{SparKV}, an overhead-adaptive KV cache loading scheme that combines cloud streaming with on-device computation to reduce TTFT for LLM inference on edge platforms. SparKV consists of three components:
\begin{itemize}[leftmargin=10pt]
\item \textbf{KV Chunk Scheduler.} The cloud partitions KV caches into indexed chunks along token, attention head, and Transformer layer dimension. The scheduler then makes a dependency-aware loading decision for each chunk, namely whether to stream a precomputed KV block or compute it locally, to minimize end-to-end TTFT.

\item \textbf{Overhead Model.} A lightweight multilayer perceptron (MLP) predictor estimates per-chunk computation latency from attention sparsity features.

\item \textbf{Runtime Controller.} An adaptive controller monitors wireless throughput and edge compute headroom within sliding windows \and dynamically migrates chunks between streaming and computation paths as conditions change.
\end{itemize}

Extensive experiments across multiple LLMs, datasets, and edge devices show that SparKV reduces TTFT by \(1.3\times\) to \(5.1\times\) compared with prior efficient KV loading schemes while maintaining response quality. Additionally, SparKV reduces per-request energy consumption by 1.5$\times$-3.3$\times$. SparKV also remains robust under varying compute availability and real wireless conditions, demonstrating its practicality for real-world deployment.

\section{Background}
\label{sec:background}

\subsection{On-device LLM Inference with Context Reuse}

Deploying on-device LLMs, typically with 0.5--7B parameters and 4-bit quantization~\cite{icml24awq}, enables privacy-preserving and low-latency inference without relying on cloud connectivity~\cite{yin2024llm}. Recent advances in quantization, pruning, and distillation have made billion-parameter models increasingly practical on edge platforms such as mobile GPUs and neural processing units (NPUs). However, on-device inference remains fundamentally constrained by limited memory bandwidth and compute throughput.

A common practice in edge LLM serving is \textit{context reuse}, where reusable context is appended to the current prompt to improve response quality and consistency. This pattern frequently arises in applications such as multi-turn conversation histories in chat assistants, private or enterprise documents for question answering, UI interaction traces for agentic tasks, and retrieved passages or media content in RAG-style applications \cite{liu2024cachegen}. In such workloads, the dominant latency often comes from constructing the corresponding KV cache before decoding can begin. This step can be performed either by recomputing the context locally during prefill phase or by loading a precomputed KV cache from external storage.

In the Transformer architecture~\cite{nips17attention}, each input token is projected into Query ($Q$), Key ($K$), and Value ($V$) vectors. Self-attention then uses the interaction between $Q$ and $K$ to compute relevance scores, which weight $V$ to produce contextualized representations. Inference consists of two phases:

\begin{itemize}[leftmargin=10pt]
    \item \textbf{Prefill Phase.} The model processes the full input context and computes the $K$ and $V$ vectors for every token across all layers, which together form the KV cache. For a model with hidden dimension $d_{\text{model}}$ and context length $L$, each layer stores $2 \times L \times d_{\text{model}}$ elements. The quadratic attention cost, $\mathcal{O}(L^2 \cdot d_{\text{model}})$, makes prefill the dominant contributor to TTFT when reusable contexts are large.
    
    \item \textbf{Decoding Phase.} The model generates tokens autoregressively. Each new token is projected into a query that attends to the cached $K$ and $V$ vectors from all prior positions, requiring only a single row of the attention matrix and reducing the per-step cost to $\mathcal{O}(L \cdot d_{\text{model}})$.
\end{itemize}

\subsection{KV Cache Loading}

For context-reuse workloads, on-device LLM serving is often bottlenecked by TTFT because edge devices have limited compute capacity for local prefill and limited memory for storing large reusable KV caches. Two complementary mechanisms can reduce this overhead.

\noindent
\textbf{KV Streaming.}
Compression techniques such as quantization and low-rank approximation~\cite{liu2024kivi,liu2024cachegen} allow precomputed KV caches to be stored remotely and streamed to the edge device, thereby bypassing local prefill. This approach is attractive for edge deployment because network transfer is often substantially more energy-efficient than on-device GPU execution; for example, driving a network interface typically consumes only 2--3\,W, compared with 20--30\,W for GPU computation on platforms such as the NVIDIA Jetson Orin NX. Moreover, modern wireless links with CDN support can deliver KV chunks at high rates. However, streaming alone cannot reliably meet latency targets under the variability of mobile network conditions.

\noindent
\textbf{Sparse Computing.}
Instead of loading precomputed KV states, sparse attention methods reduce local prefill overhead by partitioning attention computation into blocks, estimating the importance of each block, and skipping redundant operations~\cite{xu2025xattention,zhangspargeattention,zhang2024sageattention2}. This can substantially lower TTFT while keeping data and computation entirely on-device. However, unlike streaming, local computation must meet strict Transformer dependencies: a chunk can be computed only after its token-level predecessor has provided the required historical KV states and its layer-level predecessor has been fully computed.

\section{Motivation}
\label{sec:meas}

\begin{figure*}[t]
\centering
\small
\rowcolors{2}{blue!10}{white}
\begin{tabular}{cccccccccc}
\toprule
&&&& Context & \multicolumn{2}{c}{KV streaming} & \multicolumn{2}{c}{KV computing} \\
Device & GPU/NPU & NIC & Model & length & TTFT (s) & Energy (J) & TTFT (s) & Energy (J)\\
\midrule
Redmi K80 Pro & Hexagon & Wi-Fi 7 NIC &Qwen3-4B & 8K  & 2.1 & 4.8  & 13.6 & 38.7 \\
HP laptop     & RTX 5080 & Intel BE200 & Qwen3-4B     & 12K & 2.6 & 5.7  & 3.4 & 379.8\\
Jetson Orin   & Ampere 16GB & Intel AX200 & Qwen3-VL-8B & 16K & 3.5 & 11.6 & 10.9  & 354.7\\
Jetson AGX    & Ampere 64GB & Intel AX200 & Llama-3.1-8B & 24K & 6.1 & 20.8 & 13.3 & 583.6\\
\bottomrule
\end{tabular}
\vspace{-0.05in}
\captionof{table}{\small TTFT and energy comparison between wireless KV streaming and on-device prefill across representative edge platforms.}
\label{Tab:power}
\vspace{-0.2in}
\end{figure*}

This section presents a measurement study that motivates the design of SparKV. We examine three aspects of context-reuse inference on edge devices: the relative benefits of wireless KV streaming and local prefill, the heterogeneity of chunk-level overhead, and the effectiveness of naively overlapping streaming and computation. The results lead to three observations that directly guide the design of SparKV.

\noindent
\textbf{Experimental setup.}
We evaluate context loading by fetching precomputed KV caches from Aliyun~\cite{alibaba} over a wireless last-hop connection. The access point is connected to the public Internet through Gigabit Ethernet, so the dominant performance variability comes from the wireless edge link rather than the wired backhaul. Under this setting, the average cloud-to-device throughput is 850\,Mbps, with a standard deviation of 264\,Mbps. We evaluate Qwen3-4B~\cite{qwen3technicalreport}, Llama-3.1-8B~\cite{meta2024llama318b}, and Qwen3-VL-8B~\cite{qwen3technicalreport} on GPU and NPU platforms. All models are quantized to 4-bit. On GPU platforms, we run them with Hugging Face Transformers~\cite{wolf2020transformers}; on the mobile NPU platform, we use \texttt{llama.cpp}~\cite{llamacpp}. We benchmark TTFT on TriviaQA~\cite{joshi2017triviaqa}, HotpotQA~\cite{yang2018hotpotqa}, and VideoMME~\cite{fu2025video} by comparing wireless KV streaming with local prefill accelerated by SpargeAttention~\cite{zhangspargeattention}. To quantify the energy of the NIC, we utilize a Xiaomi smart plug, reporting the average power across ten trials. For the smartphone, we estimate the average NPU power by isolating the active power: we measure the total device power during NPU-exclusive inference and subtract the baseline idle power.

\subsection{Streaming Versus On-device Prefill}
\label{subsec:feasibility}

\noindent
\textbf{Observation 1: wireless KV streaming and local prefill are both viable, but they offer different tradeoffs.}
Table~\ref{Tab:power} compares TTFT and energy consumption between wireless KV streaming and on-device prefill across multiple edge platforms. Overall, wireless KV streaming consistently achieves lower TTFT and substantially lower energy consumption than pure local prefill, and its relative advantage generally becomes more pronounced as the reusable context grows. For example, wireless streaming reduces TTFT and energy by 2.2$\times$ and 28$\times$ when processing 24K-context in Jetson AGX. This trend arises because the overhead of local prefill increases super-linearly with context length, whereas KV transfer scales more gracefully with the amount of reusable cache data.

Streaming, however, is not universally preferable. Its effectiveness depends on wireless link quality, introduces cloud storage and serving overheads, and may expose privacy-sensitive context data. In contrast, local prefill preserves privacy, avoids reliance on the cloud, and can directly benefit from sparse-attention acceleration techniques~\cite{nips24minference,zhangspargeattention,xu2025xattention}. These complementary properties indicate that neither pure streaming nor pure local prefill is optimal across all deployments. Instead, an effective edge system should integrate both paths and dynamically balance latency, energy, and privacy.

\subsection{Chunk-level Overhead Heterogeneity}
\label{subsec:sparsity}

\begin{figure}
    \centering
    \includegraphics[width=0.99\linewidth]{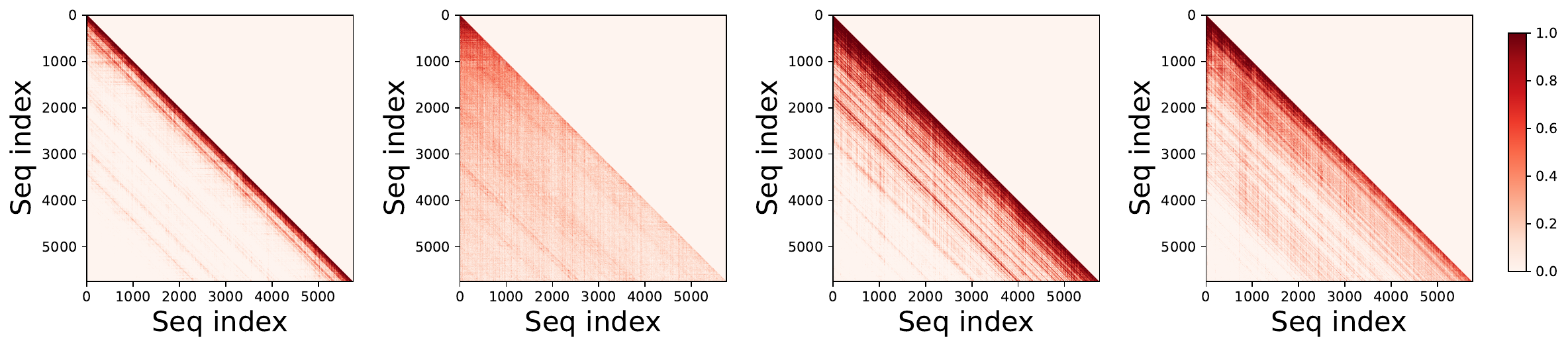}
    \includegraphics[width=0.99\linewidth]{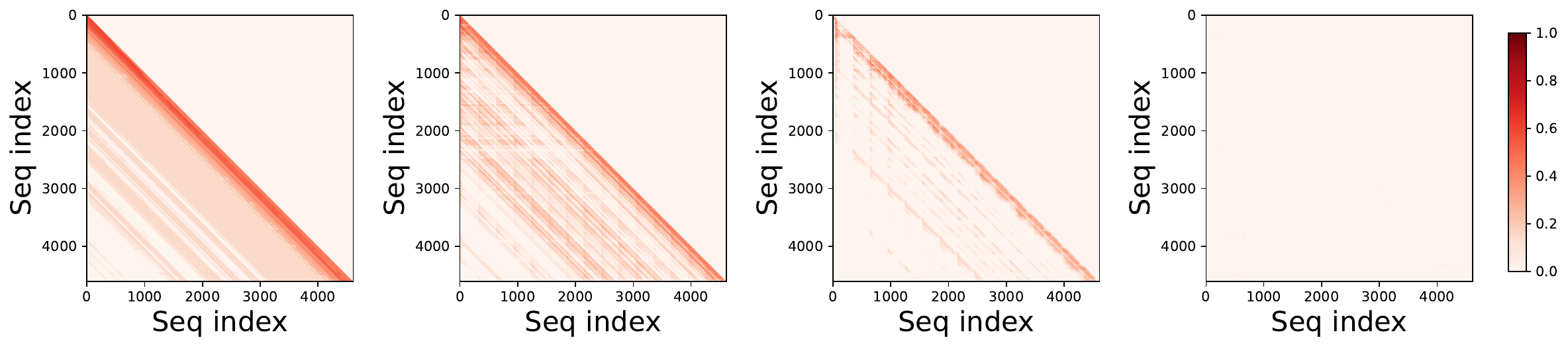}
    \vspace{-0.2in}
    \caption{\small Visualization of attention sparsity across four representative heads from Qwen3-4B (upper row) and Qwen3-VL-8B (lower row).}
    \vspace{-0.15in}
    \label{fig:att}
\end{figure}

\begin{figure}
    \centering
    \includegraphics[width=0.99\linewidth]{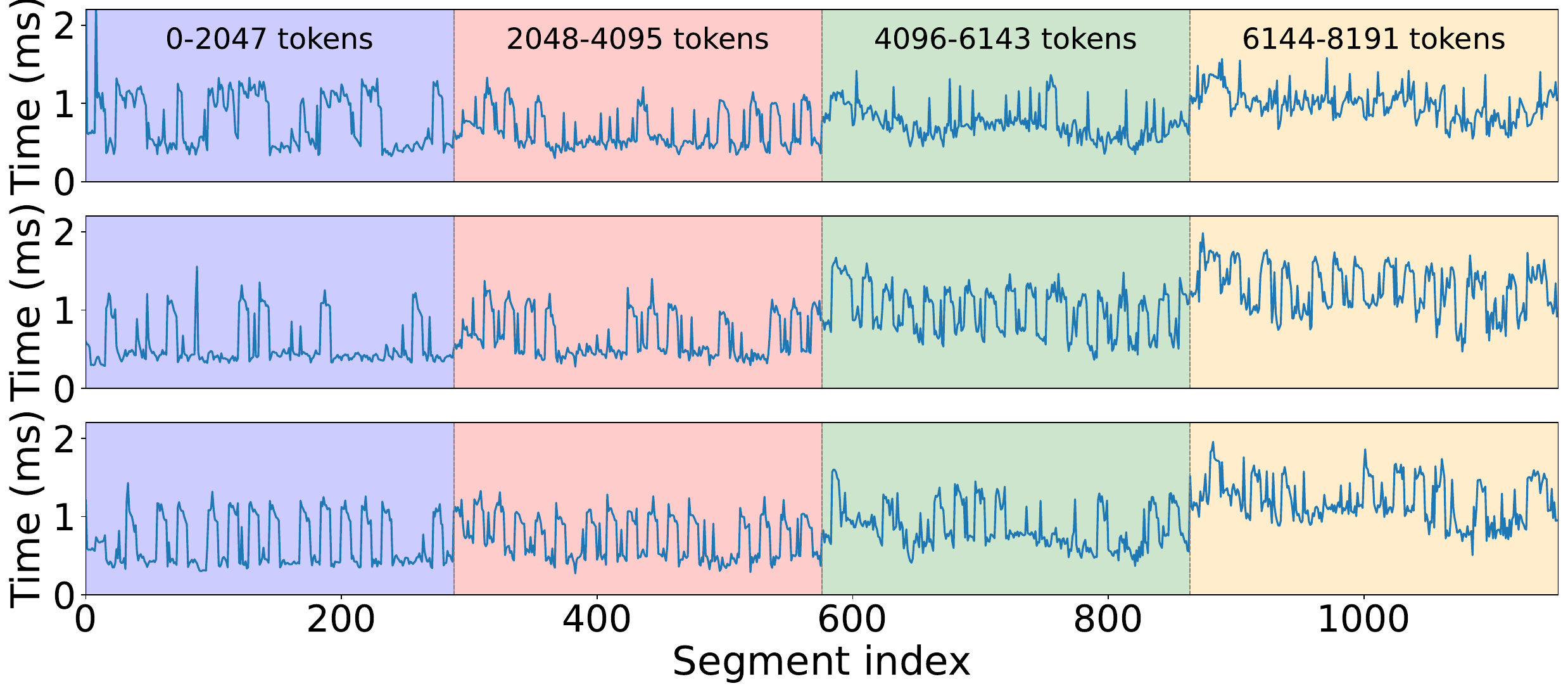}
    \vspace{-0.25in}
    \caption{\small Chunk-level computation latency of sparse attention for three samples from TriviaQA.}
    \vspace{-0.22in}
    \label{fig:att-comp}
\end{figure}

\noindent
\textbf{Observation 2: chunk-level overheads are highly heterogeneous.}
To determine whether hybrid KV loading requires fine-grained scheduling, we measure chunk-level overheads on both the computation and streaming paths.

\noindent
\textit{Computation overhead.}
We first examine the sparsity patterns of attention maps for both text QA in TriviaQA \cite{joshi2017triviaqa} and video understanding in VideoMME \cite{fu2025video}. As shown in Fig.~\ref{fig:att}, different attention heads exhibit substantially different sparsity structures, including diagonal and block-like patterns, consistent with prior observations~\cite{nips24minference}. Because sparse-attention latency depends strongly on these patterns, KV chunks can incur highly heterogeneous local prefill overheads. Using SpargeAttention~\cite{zhangspargeattention}, we partition the KV cache into 1024-token chunks across heads and layers and profile the corresponding sparse-attention latency. Fig.~\ref{fig:att-comp} shows that the compute time ranges from 0.13\,ms to 2.3\,ms, corresponding to a \(17.7\times\) variation across chunks.

\begin{figure}[t]
    \centering
    \includegraphics[width=0.85\linewidth]{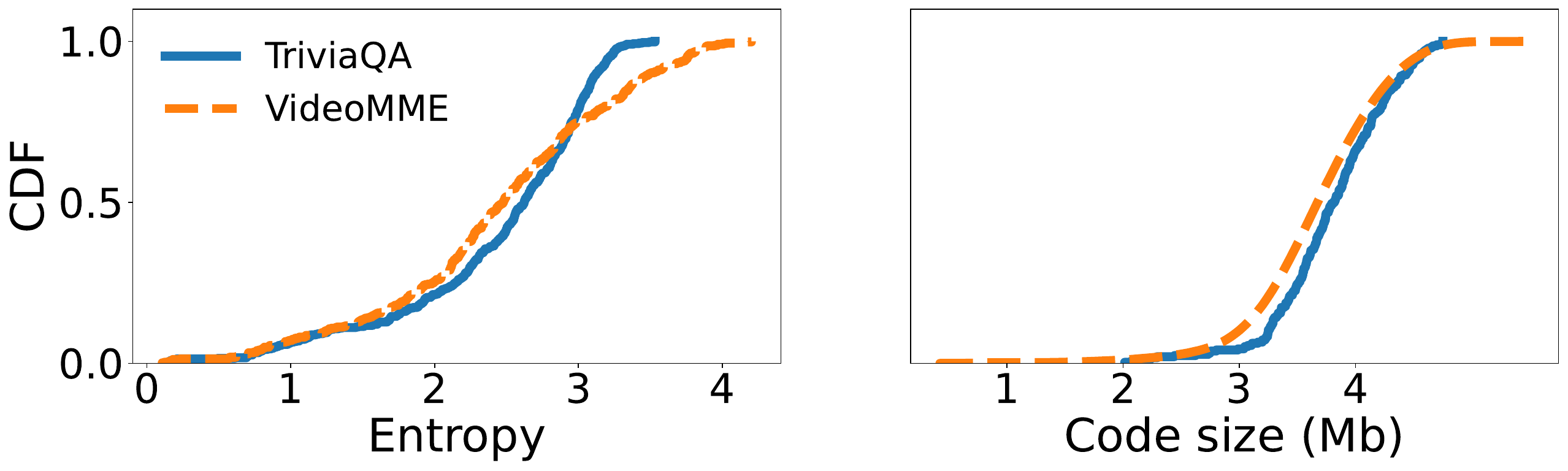}
    \vspace{-0.1in}
    \caption{\small Distribution of entropy and code size of KV cache chunks in Qwen3-4B on TriviaQA and VideoMME.}
    \vspace{-0.15in}
    \label{fig:kv_entropy}
\end{figure}

\noindent
\textit{Streaming overhead.}
We next examine the communication path. We partition the KV cache into 1024-token chunks across heads and layers, apply uniform 5-bit quantization to keys and values, and further compress them with Huffman coding~\cite{knuth1985dynamic}. As shown in Fig.~\ref{fig:kv_entropy}, the entropy varies from 0 to 4 bits per value, which leads to substantial variation in compressed KV size. Some heads compress to below 3.5\,Mb, whereas others remain much larger. Consequently, the streaming overhead also varies considerably across chunks.

\begin{figure}[t]
    \centering
    \includegraphics[width=0.99\linewidth]{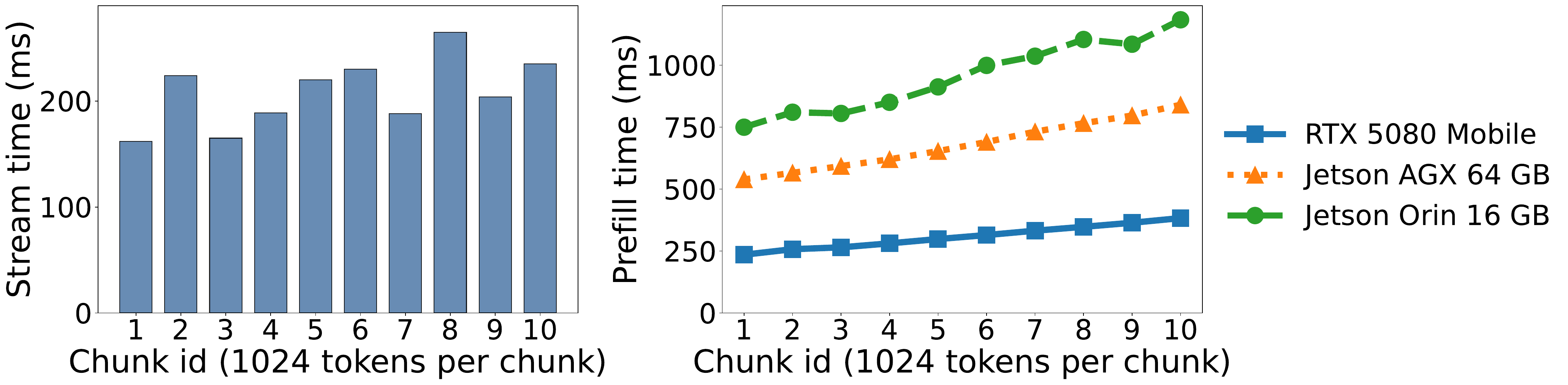}
    \vspace{-0.15in}
    \caption{\small Chunk-level streaming and local computation overhead on edge devices.}
    \vspace{-0.25in}
    \label{fig:mea_limit}
\end{figure}

\subsection{Why Naive Overlap Is Not Enough}

\noindent
\textbf{Observation 3: naively overlapping streaming and computation is insufficient in wireless edge settings.}
To evaluate whether simple overlap already captures most of the available benefit, we implement a strong hybrid baseline based on~\cite{computeorload}, augmented with Huffman compression for streaming and SpargeAttention-based local prefill computation. This pipeline streams KV chunks from Aliyun to a Jetson Orin (16\,GB) while performing local prefill in parallel. We compare it against three single-path baselines that rely exclusively on either compressed KV streaming (CacheGen~\cite{liu2024cachegen}, KIVI~\cite{liu2024kivi}) or sparse local computation (SpargeAttention~\cite{zhangspargeattention}), all under comparable response quality (F1 score $\geq$ 0.9) on TriviaQA.

The hybrid baseline outperforms the single-path baselines by \(1.4\times\) to \(1.8\times\), confirming that overlapping communication and computation is fundamentally beneficial. However, this improvement still falls short of the \(2.2\times\) speedup reported in wired server environments~\cite{computeorload}. We identify two main reasons.

\noindent
\textbf{(1) Wireless instability.}
In edge settings, throughput fluctuates over time, which disrupts the intended overlap between streaming and computation and can lead to TTFT spikes. Because the two paths progress at different and time-varying rates, a static assignment of KV chunks cannot consistently maintain high overlap.

\noindent
\textbf{(2) Chunk heterogeneity and sparsity unawareness.}
Existing hybrid schemes typically overlap earlier computation with later streaming in a fixed positional order, without accounting for the heterogeneous overhead of individual chunks. However, Fig.~\ref{fig:mea_limit} shows that chunk-level streaming overhead varies significantly over wireless links, and \cref{subsec:sparsity} further shows that sparse-attention compute time can differ by \(3\times\) to \(5\times\) even for chunks at similar positions. Positional order is therefore a poor proxy for chunk overhead; the dominant factor is the heterogeneity induced by compression behavior and attention sparsity.

These findings directly motivate SparKV. To fully exploit the complementary strengths of wireless KV streaming and on-device prefill, an edge system must be both \emph{dependency-aware} and \emph{overhead-aware}: it should schedule chunks based on their individual streaming and computation overheads, while dynamically adapting to runtime fluctuations in network throughput and edge compute availability.

\section{Design of SparKV}
\label{sec:des}

\subsection{Overview}
\label{subsec:overview}

\begin{figure}[t]
    \centering
    \includegraphics[width=0.99\linewidth]{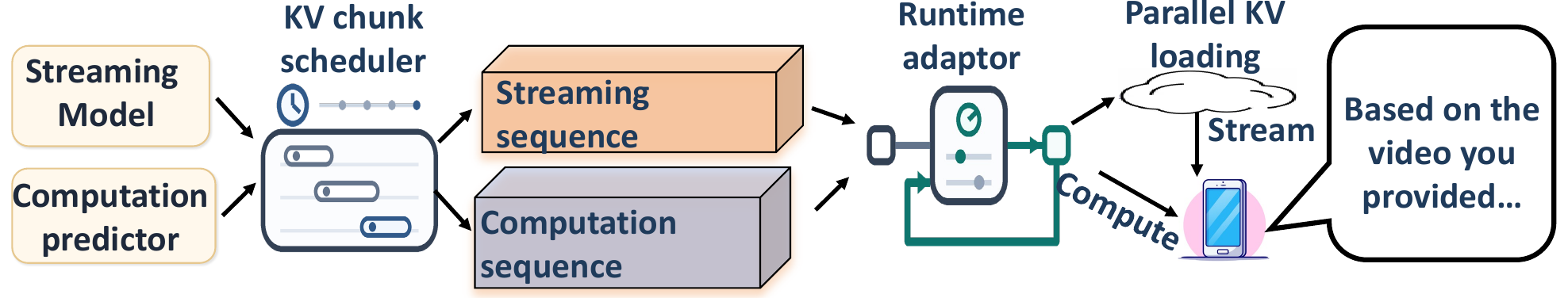}
    \vspace{-0.2in}
    \caption{\small High-level architecture of SparKV.}
    \label{fig:arch}
    \vspace{-0.2in}
\end{figure}

Fig.~\ref{fig:arch} illustrates the architecture of SparKV. The server profiles streaming overhead and trains a lightweight MLP to predict the local computation latency of each chunk. SparKV precomputes two sequences for KV streaming and local computation through the scheduler. During inference, the cloud identifies the chunks assigned to the streaming sequence and delivers them to the edge device, while the edge computes the remaining chunks locally. To accommodate runtime fluctuations in wireless throughput and edge compute availability, SparKV further employs an online controller that dynamically rebalances streaming and computation.

\subsection{KV Chunk Scheduler}
\label{subsec:scheduler}

\begin{figure}[t]
    \centering
    \includegraphics[width=0.99\linewidth]{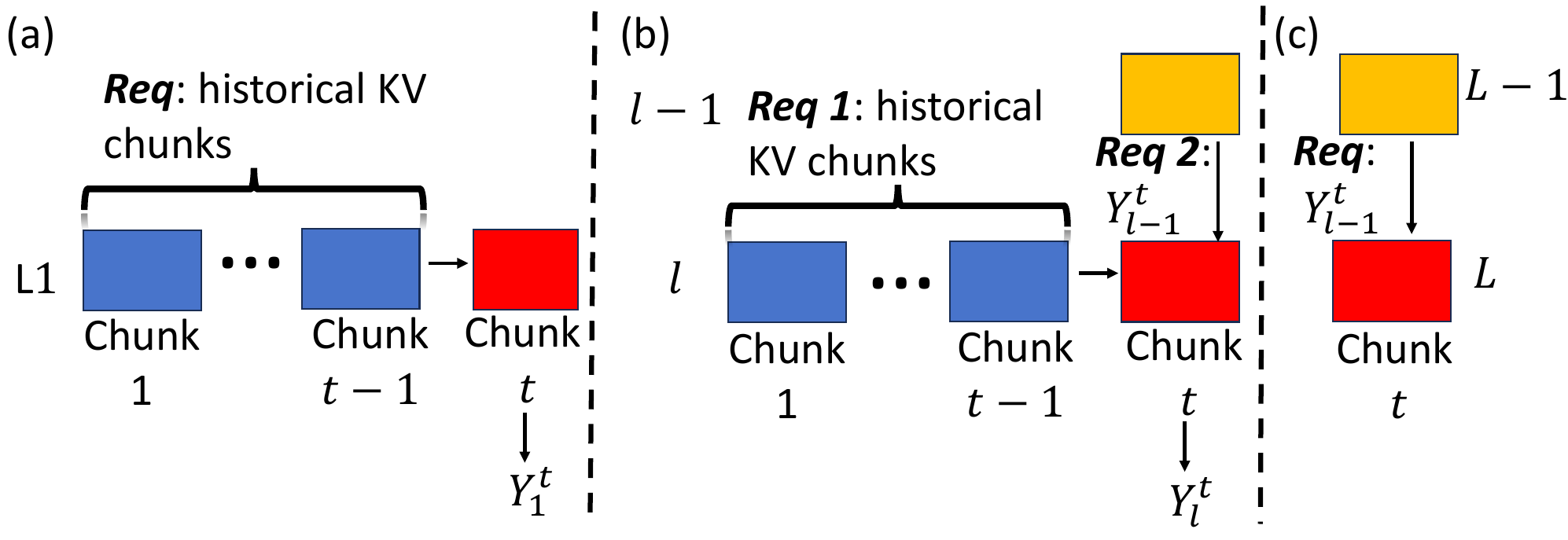}
    \vspace{-0.3in}
    \caption{\small Computation dependencies in (a) the first layer, (b) interior layers, and (c) the final layer.}
    \label{fig:schedule}
    \vspace{-0.3in}
\end{figure}

The challenge of designing the scheduler arises from the dependency structure of Transformer computation. Fig.~\ref{fig:schedule} illustrates these dependencies through three cases that determine when a chunk becomes eligible for local computation.  For an interior layer \(1 < l < L\), the KV cache of token chunk \(t\), denoted by \((K_l^t, V_l^t)\), can be computed only after chunk \(t\) has computed layer \(l-1\), since it is projected from the hidden state \(Y_{l-1}^t\). Moreover, computing layer \(l\) requires causal attention over all preceding token chunks at the same layer, namely \((K_l^{<t}, V_l^{<t})\). Therefore, an interior-layer chunk becomes eligible for computation only when both vertical dependencies across layers and horizontal dependencies across token chunks are satisfied. The boundary cases are simpler: at \(l=1\), only horizontal dependencies remain because there is no lower layer; at \(l=L\), only vertical dependencies remain because computing \((K_L^t, V_L^t)\) requires only \(Y_{L-1}^t\), and does not depend on the historical KV chunks of layer \(L\).

\noindent
\textbf{Problem formulation.}
We formulate dependency-aware chunk scheduling as a mixed-integer linear program (MILP). The KV cache is partitioned into 1024-token context chunks, which serve as the basic scheduling units. Each chunk is indexed as \(c=(t,l,h)\in\mathcal{C}\), where \(t\in[1,\lceil T/1024\rceil]\) denotes the token-chunk index, \(l\in[1,L]\) the layer index, and \(h\in[1,H]\) the attention-head index. The schedule proceeds over \(K\) decision stages. At stage \(k\), chunk \(c\) can either be streamed, computed locally, or left pending, represented by binary variables \(x_c^{\mathrm{trans}}(k)\) and \(x_c^{\mathrm{comp}}(k)\). We further introduce a binary readiness variable \(z_c(k)\), where \(z_c(k)=1\) indicates that chunk \(c\) is eligible for local computation at stage \(k\). Initially, only the first token chunk in the first layer is compute-ready, that is, \(z_{1,1,h}(1)=1\), while all other chunks are not ready at the first stage.

\noindent
\textbf{Objective function.}
Because KV streaming and local computation can overlap within a stage, the duration of stage \(k\) is determined by the slower of the two paths. The objective is therefore to minimize the total makespan:
\begin{equation}
\begin{aligned}
\min \sum_{k=1}^{K} \max\Bigl\{&
\sum_{c\in\mathcal{C}} t_{\mathrm{stream}}(c)\,x_c^{\mathrm{trans}}(k), \\
&
\sum_{c\in\mathcal{C}} t_{\mathrm{comp}}(c)\,x_c^{\mathrm{comp}}(k)
\Bigr\}.
\end{aligned}
\end{equation}
Here, \(t_{\mathrm{stream}}(c)\) denotes the estimated streaming latency of chunk \(c\), computed as
\(t_{\mathrm{stream}}(c)=b_c/\overline{bw}+t_{\mathrm{proc}}\),
where \(b_c\) is the compressed chunk size, \(\overline{bw}\) is the average download throughput profiled from ten offline trials, and \(t_{\mathrm{proc}}\) is the post-reception decryption and decoding overhead. \(t_{\mathrm{comp}}(c)\) denotes the local computation latency predicted by the model in \cref{sec:comp}.

\noindent
\textbf{Constraints.}
Each chunk must be processed exactly once, and local computation is allowed only when the chunk is compute-ready:
\begin{align}
\sum_{k=1}^{K}\Bigl(x_c^{\mathrm{trans}}(k)+x_c^{\mathrm{comp}}(k)\Bigr) &= 1,
&& \forall c\in\mathcal{C}, \\
x_c^{\mathrm{comp}}(k) &\le z_c(k),
&& \forall c\in\mathcal{C},\ \forall k.
\end{align}

To capture the computation dependencies illustrated in Fig.~\ref{fig:schedule}, we define two cumulative indicators. Token dependency is satisfied once the preceding token chunk has been streamed or computed, whereas layer dependency is satisfied only after the corresponding chunk in the previous layer has been computed locally:

{\footnotesize
\begin{align}
r^{\mathrm{tok}}_{t,l,h}(k) &=
\begin{cases}
1, & t=1 \ \text{or}\ l=L,\\
\sum_{k'=1}^{k}\Bigl(x_{t-1,l,h}^{\mathrm{trans}}(k')+x_{t-1,l,h}^{\mathrm{comp}}(k')\Bigr), & t>1 \ \text{and}\ l<L,
\end{cases} \\
r^{\mathrm{lay}}_{t,l,h}(k) &=
\begin{cases}
1, & l=1,\\
\sum_{k'=1}^{k} x_{t,l-1,h}^{\mathrm{comp}}(k'), & l>1.
\end{cases}
\end{align}
}
A chunk becomes compute-ready only when both required dependencies are satisfied.

\begin{table}[t]
\centering
\scriptsize
\setlength{\tabcolsep}{4pt}
\begin{tabular}{@{}llcccc@{}}
\toprule
\multirow{2}{*}{\textbf{Dataset}} & \multirow{2}{*}{\textbf{Len.}} & \multicolumn{2}{c}{\textbf{Greedy}} & \multicolumn{2}{c}{\textbf{Gurobi}} \\
\cmidrule(lr){3-4} \cmidrule(lr){5-6}
& & \textbf{Runtime (s)} & \textbf{TTFT (s)} & \textbf{Runtime (s)} & \textbf{TTFT (s)} \\
\midrule
\multirow{2}{*}{LongChat}
& 10K & 10.7 & 1.12 & 122.5 & 1.08 \\
& 20K & 22.3 & 2.31 & 673.4 & 2.25 \\
\midrule
\multirow{2}{*}{VideoMME}
& 10K & 12.3 & 1.47 & 114.1 & 1.41 \\
& 20K & 25.4 & 2.39 & 705.3 & 2.33 \\
\bottomrule
\end{tabular}
\vspace{-0.05in}
\caption{\small Comparison between the proposed greedy heuristic and exact MILP solving with Gurobi.}
\label{tab:algorithm_comparison}
\vspace{-0.3in}
\end{table}

\noindent
\textbf{Potential-aware greedy heuristic.}
Although the MILP provides an oracle formulation, solving it exactly is too expensive for practical deployment because the search space grows rapidly with the number of chunks and decision stages. Moreover, a naive latency-only greedy policy is insufficient: the benefit of scheduling a chunk depends not only on its own overhead, but also on the future computation opportunities it unlocks. SparKV therefore adopts a \textit{potential-aware greedy heuristic} that approximates the MILP using lightweight priority scores.

For each chunk \(c\), SparKV assigns a streaming priority \(w_s(c)\) and a computation priority \(w_c(c)\), where a larger value indicates higher priority. Both scores combine the immediate overhead of the current chunk with the potential benefit of newly enabled computation:
\[
\begin{aligned}
w_s(c) &= \frac{1}{t_{\mathrm{stream}}(c)}
+ \sum_{c'\in\mathcal{A}_s(c)} \frac{1}{t_{\mathrm{comp}}(c')}, \\
w_c(c) &= \frac{1}{t_{\mathrm{comp}}(c)}
+ \sum_{c'\in\mathcal{A}_c(c)} \frac{1}{t_{\mathrm{comp}}(c')}.
\end{aligned}
\]
Here, \(\mathcal{A}_s(c)\) and \(\mathcal{A}_c(c)\) denote the sets of chunks that become newly compute-ready after streaming or locally computing chunk \(c\), respectively. Intuitively, \(w_s\) favors chunks that are inexpensive to stream and can unlock additional low-overhead local computation, whereas \(w_c\) favors chunks that are inexpensive to compute and can further advance the computation frontier. By default, SparKV assigns equal weights to these terms, although the relative weights can be adjusted at deployment time to trade off privacy, energy consumption, and TTFT.

At each stage \(k\), SparKV maintains two queues: a computation sequence \(Q_c\) containing compute-ready chunks awaiting local execution and a streaming sequence \(Q_s\) containing chunks not yet loaded. The scheduler first sorts \(Q_c\) and \(Q_s\) in descending order of \(w_c\) and \(w_s\), respectively. It then greedily schedules local computation under a time budget \(\Delta t\): the highest-priority chunk is removed from \(Q_c\), scheduled for execution, and charged a overhead of \(t_{\mathrm{comp}}(c)\). Once a chunk is selected for local computation, it is removed from \(Q_s\), because it no longer needs to be streamed. Because local computation may unlock additional chunks, \(Q_c\) is updated and re-sorted after each selection. After the computation phase ends, SparKV resets the budget to \(\Delta t\) and greedily schedules streaming from \(Q_s\) in descending order of \(w_s\). The queues are then updated for stage \(k+1\) according to the chunks newly activated by the selected operations.

\noindent
\textbf{Comparison with exact MILP solving.}
We compare the proposed heuristic with Gurobi~\cite{gurobi}, which solves the MILP formulation as a standard exact baseline. As shown in Table~\ref{tab:algorithm_comparison}, the heuristic incurs much lower scheduling overhead while achieving similar TTFT on LongChat. For a 10K-token context, the heuristic reduces scheduling runtime by \(11.4\times\), and the advantage increases to \(30.2\times\) at 20K tokens. This widening gap reflects the poor scalability of exact MILP solving, whereas the proposed heuristic requires only stage-local sorting and queue updates, making it well suited for latency-sensitive edge inference.

\subsection{Computation Latency Predictor}
\label{sec:comp}

Accurate estimation of local chunk latency, denoted by \(t_{\mathrm{comp}}(t,l,h)\), is critical for scheduling. We first decompose computation latency across layers. The final layer is a boundary case: for a chunk at layer \(L\), generating the KV cache requires only projecting the hidden states from the previous layer, so the latency reduces to a lightweight projection overhead, \(t_{\mathrm{comp}}(t,L,h)=t_{\mathrm{proj}}\). For all non-final layers \(1 \le l < L\), computing the KV cache requires executing the full Transformer layer so that the dependency chain can continue upward. We therefore decompose the chunk latency as
\[
t_{\mathrm{comp}}(t,l,h)=t_{\mathrm{attn}}(t,l,h)+t_{\mathrm{dense}},
\]
where \(t_{\mathrm{attn}}(t,l,h)\) is the sparse-attention overhead and \(t_{\mathrm{dense}}=t_{\mathrm{qkv}}+t_{\mathrm{o}}+t_{\mathrm{res}}+t_{\mathrm{norm}}+t_{\mathrm{ffn}}\) aggregates the remaining dense operators, including QKV projection, output projection, residual addition, layer normalization, and feed-forward computation. Empirically, sparse attention dominates the variation in chunk latency, whereas the dense component behaves largely as a small and nearly constant offset because the chunk shape is fixed across layers and heads.

To reduce the quadratic overhead of standard attention on edge devices, SparKV adopts block-sparse attention~\cite{zhangspargeattention}. Specifically, the query sequence is partitioned into groups of 128 tokens, while the key and value sequences are divided into 64-token sub-blocks. Under this design, the latency of a non-final-layer chunk depends not only on sequence length but also strongly on the sparsity pattern of the attention mask. This motivates a lightweight predictor to estimate the sparse-attention latency efficiently at runtime.

\noindent
\textbf{Why analytical models fall short.}
Traditional analytical models such as Roofline~\cite{williams2009roofline} estimate computation time as
$
t_{\text{roofline}}=\max\!\left(\frac{W}{P_{\text{peak}}},\frac{Q}{B_{\text{peak}}}\right),
$
where \(W\) and \(Q\) denote the computational workload and memory traffic, and \(P_{\text{peak}}\) and \(B_{\text{peak}}\) denote peak compute throughput and memory bandwidth. Although lightweight, such models assume regular computation and near-ideal hardware utilization. These assumptions do not hold for block-sparse attention, whose irregular sparsity leads to non-contiguous memory accesses and poor GPU utilization that are difficult to capture analytically. As a result, static analytical models can deviate substantially from measured latency on edge GPUs.

We find that the latency variation in \(t_{\mathrm{attn}}(t,l,h)\) is primarily determined by three factors: sequence length, attention sparsity, and instantaneous device load. We therefore represent each non-final-layer chunk using a compact feature vector \(\mathbf{x}_{\mathrm{comp}}=\langle t,s,U_{\mathrm{edge}}\rangle\), where \(t\) is the token-block index, corresponding to a query length of \(1024\times t\), \(s\) is the number of active blocks in the attention mask, and \(U_{\mathrm{edge}}\) is the real-time GPU utilization measured by \texttt{nvidia-smi}. Active blocks are defined as the most significant blocks that together account for 98\% of the total attention mass. These features capture the dominant factors affecting sparse-attention latency and enable a lightweight MLP predictor to estimate chunk-level compute time at runtime. Compared with heuristic rules or offline lookup tables, the predictor adapts to runtime variations in sparsity and GPU load with negligible overhead.

\noindent
\textbf{Lightweight predictor.}
We use a hybrid estimator:
\[
\hat t_{\mathrm{comp}}(t,l,h)=
\begin{cases}
t_{\mathrm{proj}}, & l=L,\\
f_{\theta}(\mathbf{x}_{\mathrm{comp}})+t_{\mathrm{dense}}, & 1 \le l < L,
\end{cases}
\]
where \(f_{\theta}\) is a MLP~\cite{popescu2009multilayer} that predicts the dominant sparse-attention latency of non-final layers. This design captures the dominant source of latency variation while treating the remaining dense operators as a small offset. The MLP has two hidden layers with 48 and 24 neurons, respectively, which provides a good balance between prediction accuracy and runtime overhead. We train it offline on 6,000 samples using an 80/20 train-test split, stochastic gradient descent, and mean squared error loss. This one-time training process takes only 17.6\,s on a Jetson Orin 16GB.

\begin{figure}[t]
    \centering
    \includegraphics[width=0.95\linewidth]{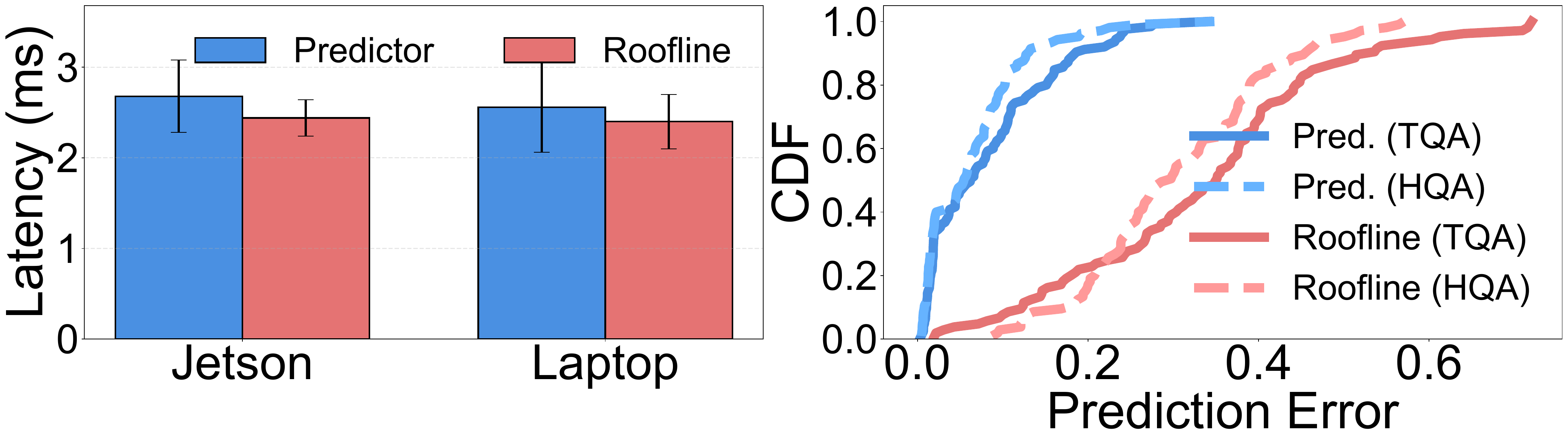}
    \vspace{-0.1in}
    \caption{\small Overhead and prediction error of the proposed predictor and the Roofline baseline for chunk computation latency estimation.}
    \vspace{-0.2in}
    \label{fig:pred}
\end{figure}

\noindent
\textbf{Overhead and accuracy.}
We compare the proposed predictor with a Roofline baseline on TriviaQA and HotpotQA using an NVIDIA Jetson Orin. For non-final-layer chunks, the predictor incurs only 2.6\,ms per-chunk inference overhead, close to the 2.0\,ms required by the Roofline baseline merely to compute \(W\) and \(Q\). The final-layer case is handled by direct profile lookup and introduces negligible additional overhead. Despite similar runtime overhead, the proposed predictor reduces latency estimation error by \(4.8\times\) to \(5.6\times\), demonstrating that learning the attention-dominated component captures the nonlinear behavior of block-sparse attention much more accurately than static analytical models.

\subsection{Runtime Adaptation Mechanism}
\label{subsec:adapt}

The schedule used by SparKV is derived from predicted streaming and computation overheads, whereas the actual execution environment is inherently dynamic. Both cloud-side KV streaming latency and local prefill latency can vary substantially because of wireless fluctuations and transient GPU contention. As a result, a schedule that is near-optimal offline may become suboptimal at runtime, reducing the overlap between communication and computation. SparKV therefore includes a runtime adaptation mechanism that adjusts the execution plan online in response to transient resource imbalance. To avoid oscillation, the controller limits the number of migrations within each stage.

\noindent
\textbf{Edge compute contention.}
When the edge GPU becomes slower than expected, local prefill falls behind KV streaming and computation becomes the transient bottleneck. If SparKV strictly follows the offline schedule, the network can become under-utilized while the GPU remains saturated. To avoid this imbalance, the runtime controller first speculatively prefetches chunks from the next scheduling stage whenever dependencies allow. If the current streaming queue is still insufficient to keep the link busy, SparKV further migrates a portion of the workload originally assigned to local computation to the streaming path. To minimize interference with chunks already close to execution, these migrated chunks are selected from the tail of the computation order rather than the head.

\noindent
\textbf{Wireless bandwidth volatility.}
When wireless throughput drops below the profiled level, KV streaming becomes the transient bottleneck and fewer chunks can be delivered within a stage than originally planned. The GPU may finish its assigned local prefill work early and become under-utilized. To preserve overlap, SparKV shifts work in the opposite direction: it identifies compute-ready chunks whose dependencies have already been satisfied but that were originally assigned to streaming, and executes them locally instead. If no such candidates remain in the current stage, the controller speculatively advances to compute-ready chunks in the next stage. In this way, SparKV converts otherwise idle GPU cycles into useful progress and mitigates the performance loss caused by temporary bandwidth degradation.

\section{Implementation}
\label{sec:impl}

SparKV works as a standalone layer to accelerate context loading and integrates with SpargeAttention and HuggingFace framework \cite{wolf2020transformers}.

\noindent
\textbf{Integration with Sparse Attention.} \textsc{SparKV} intercepts the execution of each decoder layer and replaced the standard attention operator with the optimized \texttt{block\_sparse\_sage2\_attn\_cuda} kernel from \textsc{SpargeAttention}~\cite{zhangspargeattention}. After generating the $Q$, $K$, and $V$ tensors through linear projections, the system dynamically derives attention masks from intermediate self-attention scores and measures the per-layer decoding latency.

\noindent
\textbf{Integration with Decoding Framework.} 
We integrate \textsc{SparKV} directly into the generation pipeline of the Hugging Face \texttt{transformers} library. Specifically, \textsc{SparKV} intercepts the inference initialization to reconstruct the context KV cache using its parallel streaming and computing mechanism. The assembled cache is then injected into the \texttt{model.generate()} interface as the \texttt{past\_key\_values} argument. To TTFT measurement, we set \texttt{max\_new\_tokens=1}, ensuring only the first response token is generated. For response quality assessment, decoding process until an end-of-sequence (EOS) token is produced or a maximum token budget is reached.

\noindent
\textbf{KV Cache Compression and Dataflow.} 
\textsc{SparKV} minimizes transmission overhead by compressing the KV cache into a bitstream using layer-wise non-uniform quantization and Huffman coding. After fetching the cloud-hosted artifacts into the edge device's system RAM, the decoding dataflow adapts to the underlying hardware. On Jetson platforms, the Unified Memory Architecture (UMA) allows the GPU to access system RAM directly without redundant copies. While on x86 platforms, the KV cache must be transferred explicitly from system RAM to the discrete VRAM via PCIe.

\section{Evaluation}
\label{sec:method}

\subsection{Experimental Setup}

\noindent
\textbf{Models and platforms.}
We evaluate \textsc{SparKV} on five state-of-the-art Transformer models spanning both text-only and multimodal workloads: three LLMs, \texttt{Qwen3-4B}~\cite{qwen3technicalreport}, \texttt{Llama-3.1-8B}~\cite{llama3herd2024}, and \texttt{Qwen3-14B}~\cite{qwen3technicalreport}, and two VLMs, \texttt{Qwen2.5-VL-7B}~\cite{qwen3technicalreport} and \texttt{InternVL2-8B}~\cite{chen2024internvl}. We deploy these models with Hugging Face Transformers~\cite{wolf2020transformers} on an RTX 5080 laptop GPU and a Jetson AGX. All models are quantized to 4-bit to match the memory constraints of edge deployment.

\begin{table}[t]
  \centering
  \footnotesize
  \rowcolors{2}{gray!25}{white}
  \begin{tabular}{lllll}
    \toprule
    Dataset & Task & Metric & Mean Len. & Samples \\
    \midrule
    RepoBench-P~\cite{liu2023repobench} & Code & Edit Sim. & 10K & 100 \\
    HotpotQA~\cite{yang2018hotpotqa} & Multi QA & F1 & 11K & 100 \\
    TriviaQA~\cite{joshi2017triviaqa} & Few-shot QA & F1 & 11K & 100 \\
    LongChat~\cite{longchat2023} & Topic & Accuracy & 12K & 100 \\
    GovReport~\cite{huang2021gov} & Summarization & Rouge-L & 13K & 100 \\
    NarrativeQA~\cite{kovcisky2018narrativeqa} & Single QA & F1 & 18K & 200 \\
    Academic~\cite{bai2023longbench} & Multi QA & Accuracy & 28K & 50 \\
    Financial~\cite{bai2024longbench} & Single QA & Accuracy & 49K & 22 \\
    VideoMME~\cite{fu2025video} & Video QA & Accuracy & 23K & 50 \\
    \bottomrule
  \end{tabular}
  \caption{\small Summary of the evaluation datasets.}
  \label{Tab:data}
  \vspace{-0.2in}
\end{table}

\noindent
\textbf{Datasets and tasks.}
We evaluate SparKV on nine public datasets drawn from LongBench~\cite{bai2023longbench,bai2024longbench} and VideoMME~\cite{fu2025video}, covering code completion, topic classification, single- and multi-document question answering, summarization, and video understanding.

\noindent
\textbf{Network environment.}
We build a Wi-Fi 6~\cite{mozaffariahrar2022wifi6} testbed in a production campus network to expose the system to realistic wireless interference. The local wireless link achieves 1.1--2.0\,Gbps throughput, with a standard deviation of 0.24--0.35\,Gbps. For KV loading, we store precomputed KV caches on Aliyun~\cite{alibaba}, yielding an average end-to-end cloud-to-edge throughput of 0.64\,Gbps.

\noindent
\textbf{Evaluation metrics.}
We report three metrics:
\begin{enumerate}[leftmargin=*, noitemsep, topsep=0pt]
    \item \textbf{TTFT:} end-to-end latency from request submission to the generation of the first output token.
    \item \textbf{Response quality:} task-specific metrics, including F1, Rouge-L, and accuracy, following the official LongBench and VideoMME evaluation protocols.
    \item \textbf{Energy per request:} end-to-end energy consumed on the edge device from request submission until the completion of response generation. 
\end{enumerate}

\noindent
\textbf{Baselines.}
We compare SparKV against three baselines.

\begin{itemize}[leftmargin=*]
\item \textbf{CacheGen}~\cite{liu2024cachegen} pre-encodes the KV cache into five bitrate levels using layer-wise quantization and arithmetic coding, and dynamically selects the bitrate according to the available bandwidth. We set its service-level objective to 2\,s, following prior interactive-application settings~\cite{ren2024industrial,li2025next}.

\item \textbf{Strong Hybrid}~\cite{computeorload} overlaps local computation of earlier KV chunks with streaming of later KV chunks using a fixed hybrid pipeline. To ensure a fair comparison, we strengthen this baseline with the same implementation primitives used in SparKV: the streamed portion uses the same KV quantization and Huffman coding pipeline, the locally recomputed portion uses the same SpargeAttention~\cite{zhangspargeattention} kernel, and both methods use the same chunk partitioning strategy.

\item \textbf{Local Prefill} computes the full KV cache locally on the edge GPU using the same SpargeAttention~\cite{zhangspargeattention} implementation as SparKV, without any KV streaming.
\end{itemize}

\subsection{Overall Performance}
\label{subsec:over}

\begin{figure*}[t]
    \centering
    \includegraphics[width=0.95\linewidth]{figures_1/Evaluation/eva_overall_laptop_ttft.pdf}
    \includegraphics[width=0.95\linewidth]{figures_1/Evaluation/eva_overall_laptop_per.pdf}
    \vspace{-0.1in}
    \caption{\small Overall TTFT and response quality across datasets on an RTX 5080 laptop GPU with Llama-3.1-8B.}
    \label{fig:lap_ttft}
    \vspace{-0.2in}
\end{figure*}

\noindent
\textbf{Performance across datasets.}
We first compare SparKV with all baselines across representative datasets on two edge platforms: an RTX 5080 laptop GPU and a Jetson AGX.

\noindent
\textit{Laptop results.}
Fig.~\ref{fig:lap_ttft} shows that SparKV consistently achieves the lowest TTFT across all evaluated datasets on the RTX 5080 laptop. Compared with \textit{Local Prefill}, SparKV reduces TTFT by \(2.9\times\) to \(5.1\times\) while preserving task quality. Compared with \textit{CacheGen}, SparKV achieves \(1.8\times\) to \(2.2\times\) lower TTFT and improves the median task metric by 0.07--0.10. This advantage is expected because streaming-only methods remain constrained by wireless throughput and often rely on more aggressive compression. The \textit{Strong Hybrid} baseline narrows the gap by overlapping computation and streaming, but it remains about \(1.3\times\) slower than SparKV. This result shows that overlap alone is insufficient; chunk-level scheduling that accounts for heterogeneous streaming and computation overheads is also necessary.

\begin{figure}[t]
    \centering
    \includegraphics[width=0.9\linewidth]{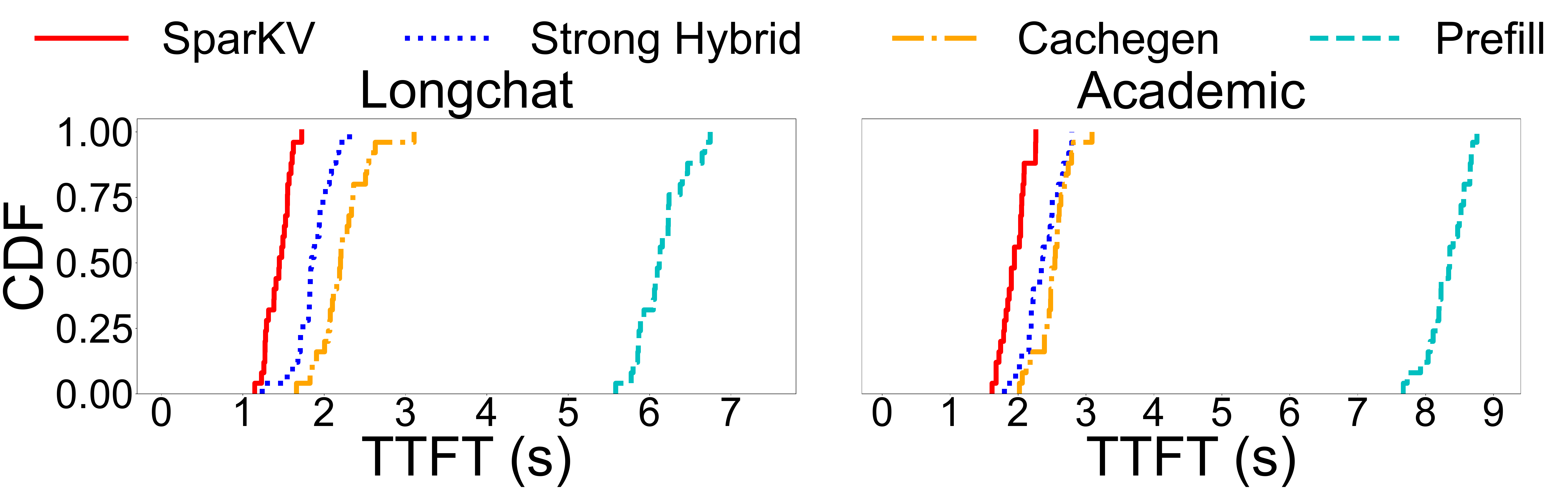}
    \includegraphics[width=0.9\linewidth]{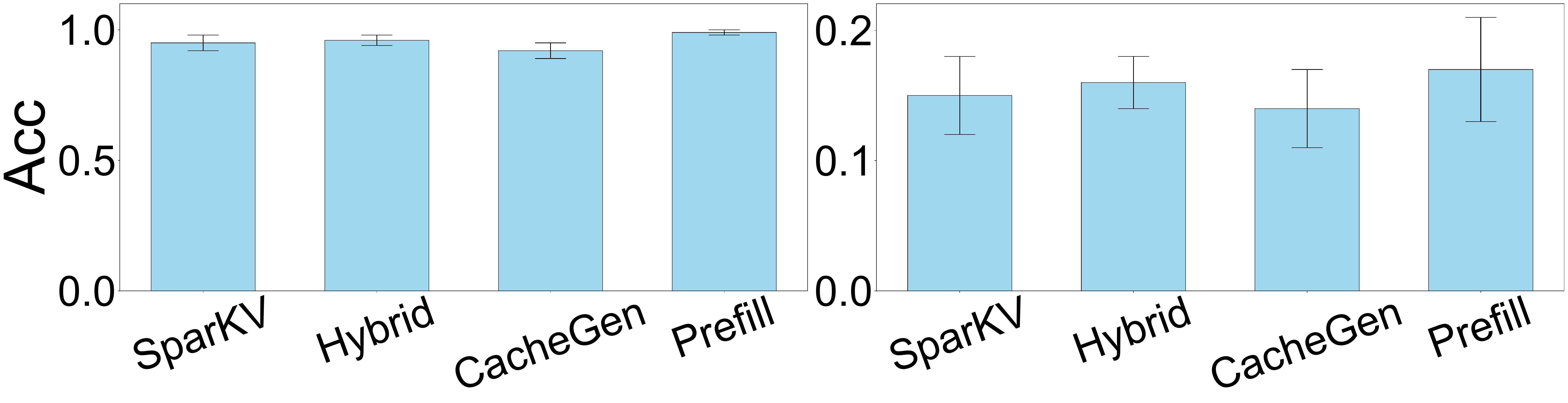}
    \vspace{-0.1in}
    \caption{\small TTFT and response quality of SparKV and baselines on a Jetson AGX 64GB with Llama-3.1-8B.}
    \vspace{-0.2in}
    \label{fig:jet_ttft}
\end{figure}

\noindent
\textit{Jetson results.}
Fig.~\ref{fig:jet_ttft} shows that SparKV also performs consistently well on the Jetson platform. Across the representative datasets, SparKV achieves TTFTs of 1.2--1.4\,s while maintaining response quality close to that of full local prefill. It outperforms \textit{Strong Hybrid}, \textit{CacheGen}, and \textit{Local Prefill} by up to \(1.3\times\), \(1.9\times\), and \(3.8\times\), respectively. These results confirm that the benefit of dependency-aware, overhead-aware scheduling carries over to lower-power edge hardware with more limited compute resources.

\noindent
\textbf{Performance across model families.}
We next evaluate SparKV across multiple LLMs and VLMs to test its robustness across model scales and modalities.

\begin{figure}[t]
    \centering
    \includegraphics[width=0.9\linewidth]{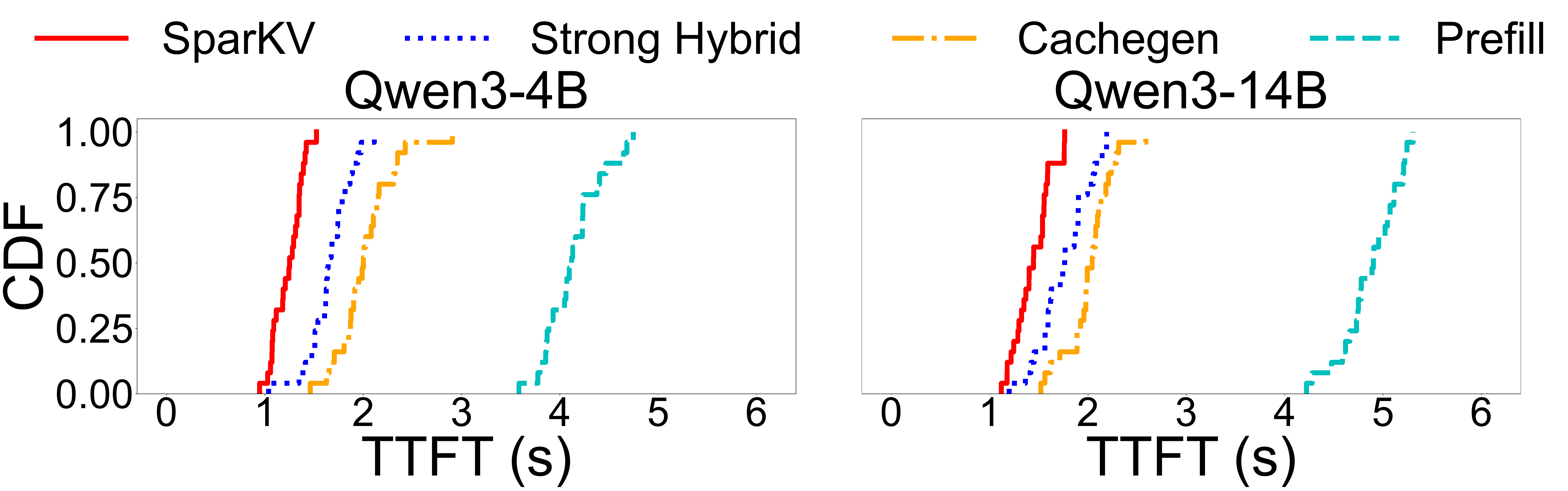}
    \includegraphics[width=0.9\linewidth]{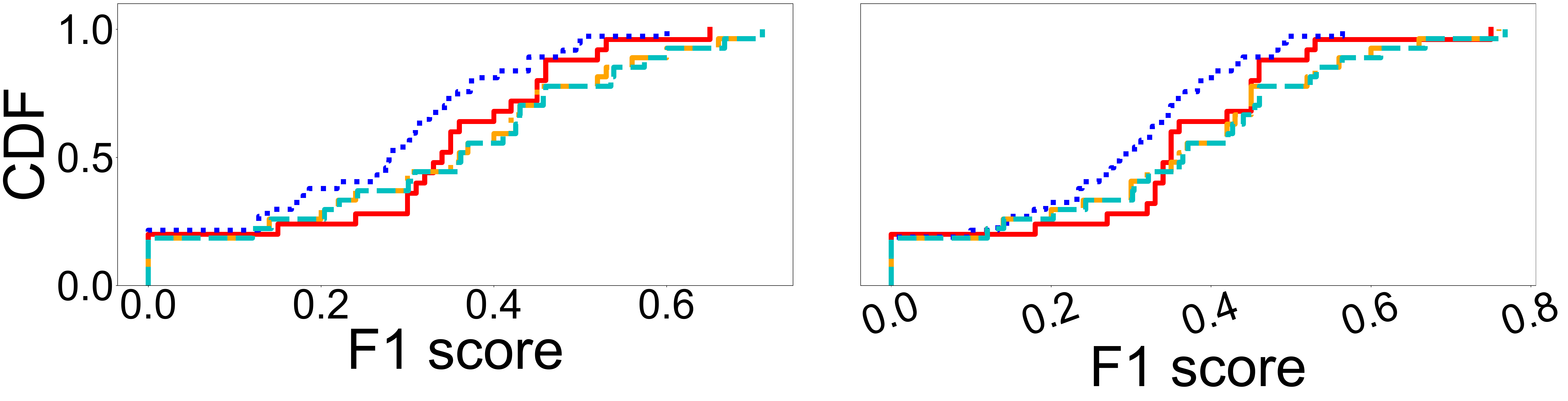}
    \vspace{-0.1in}
    \caption{\small TTFT and response quality of SparKV and baselines on HotpotQA using a laptop GPU with Qwen3-4B and Qwen3-14B.}
    \vspace{-0.2in}
    \label{fig:model_ttft_1}
\end{figure}

\noindent
\textit{LLMs.}
Fig.~\ref{fig:model_ttft_1} reports the results for Qwen3-4B and Qwen3-14B. In both cases, SparKV improves TTFT by about \(1.3\times\) over \textit{Strong Hybrid} while maintaining comparable F1 scores. This result indicates that the benefit of chunk-level scheduling is not limited to a specific model family or parameter scale.

\begin{figure}[t]
    \centering
    \includegraphics[width=0.9\linewidth]{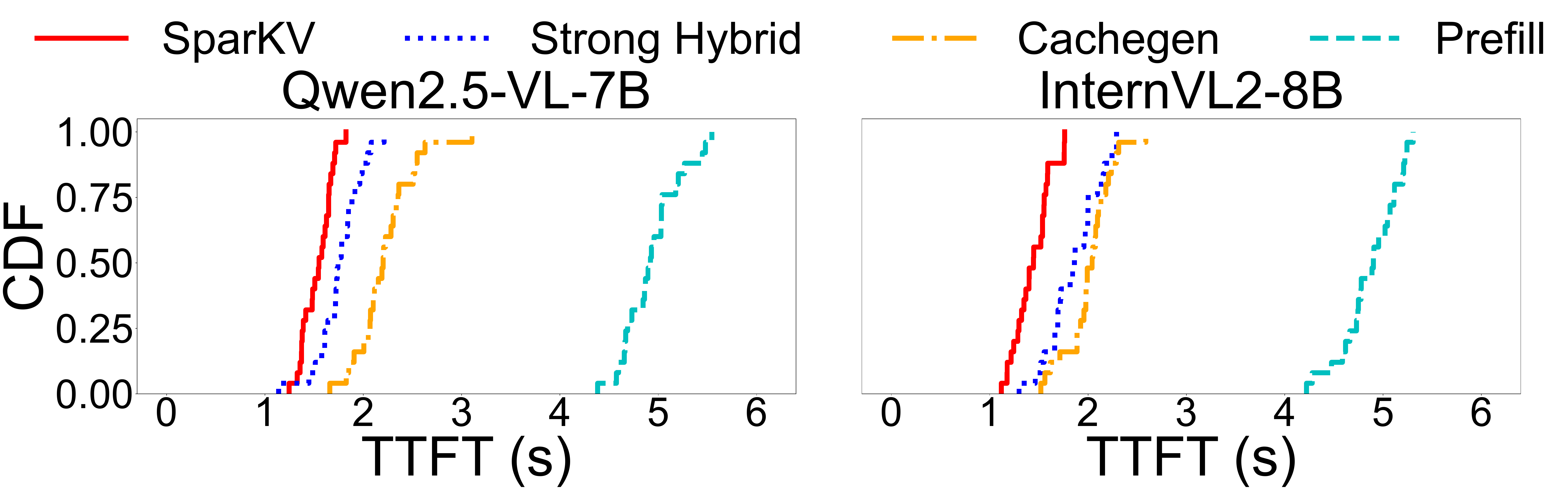}
    \includegraphics[width=0.9\linewidth]{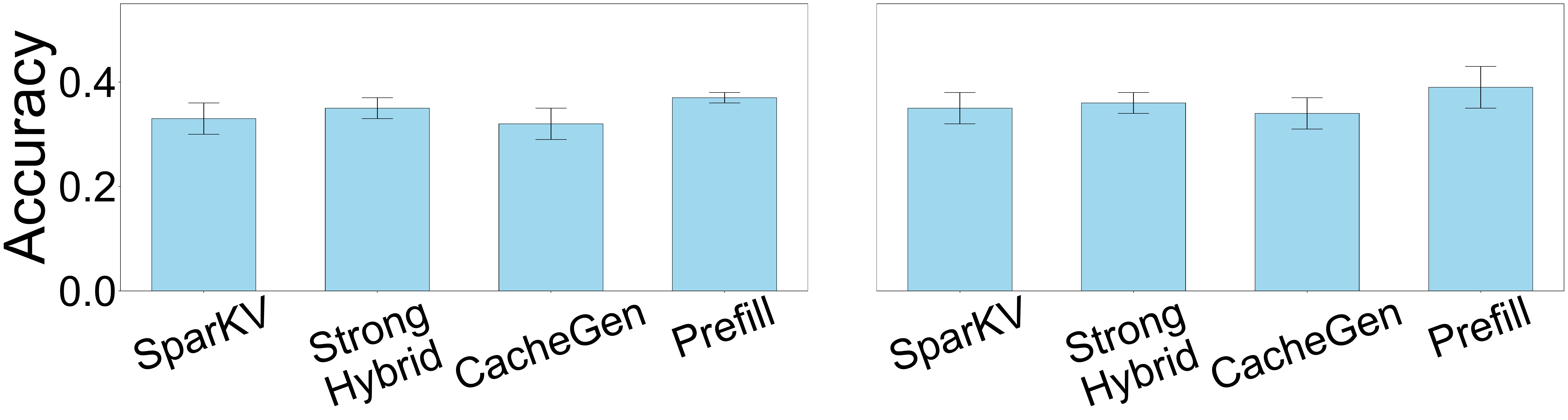}
    \vspace{-0.1in}
    \caption{\small TTFT and response quality of SparKV and baselines on VideoMME using a laptop GPU across VLMs.}
    \vspace{-0.2in}
    \label{fig:model_ttft_2}
\end{figure}

\noindent
\textit{VLMs.}
Fig.~\ref{fig:model_ttft_2} shows the results for Qwen2.5-VL-7B and InternVL2-8B on VideoMME. In this setting, SparKV delivers even larger gains, reducing TTFT by \(1.3\times\) to \(1.4\times\) relative to \textit{Strong Hybrid} and by \(1.8\times\) to \(2.0\times\) relative to \textit{CacheGen}. This larger margin is consistent with the stronger chunk-level variance in multimodal workloads, where visual tokens induce greater heterogeneity in both transmission size and attention overhead. Overall, SparKV outperforms prior schemes for two reasons. First, it avoids the single-resource bottlenecks of streaming-only and compute-only methods by overlapping the two paths. Second, it improves over prior hybrid designs through overhead-aware and dependency-aware chunk scheduling that better matches heterogeneous runtime overheads.

\subsection{Sensitivity and Robustness}
\label{subsec:bench}

\begin{figure*}[t]
    \centering
    \begin{minipage}{0.32\linewidth}
    \includegraphics[width=0.99\linewidth]{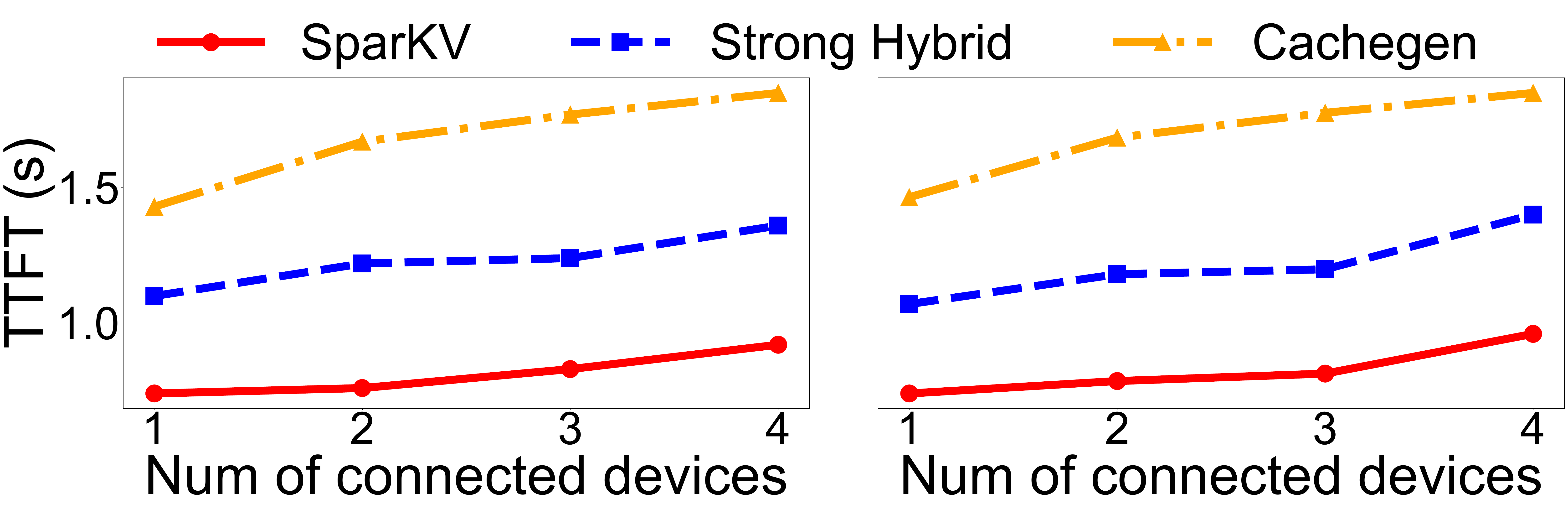}
    \vspace{-0.2in}
    \caption{\small Impact of wireless interference.}
    \label{fig:inter_ttft}
    \vspace{-0.2in}
    \end{minipage}
    \centering
    \begin{minipage}{0.32\linewidth}
        \includegraphics[width=0.99\linewidth]{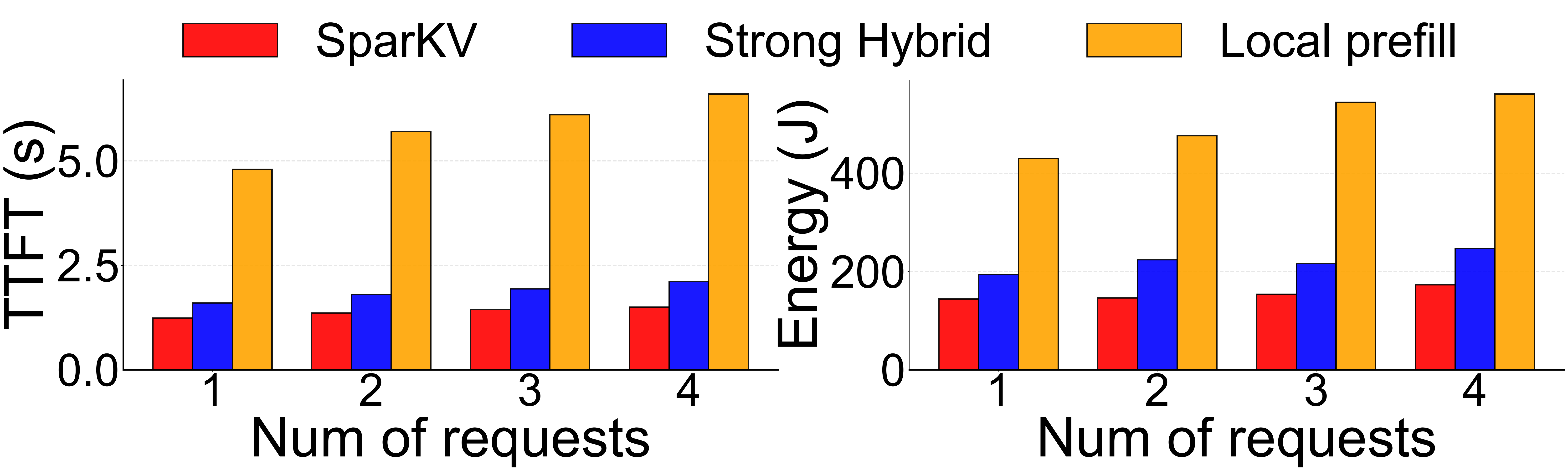}
        \vspace{-0.2in}
        \caption{\small Impact of concurrent requests.}
        \label{fig:req_ttft}
        \vspace{-0.2in}
    \end{minipage}
    \centering
    \begin{minipage}{0.32\linewidth}
        \includegraphics[width=0.99\linewidth]{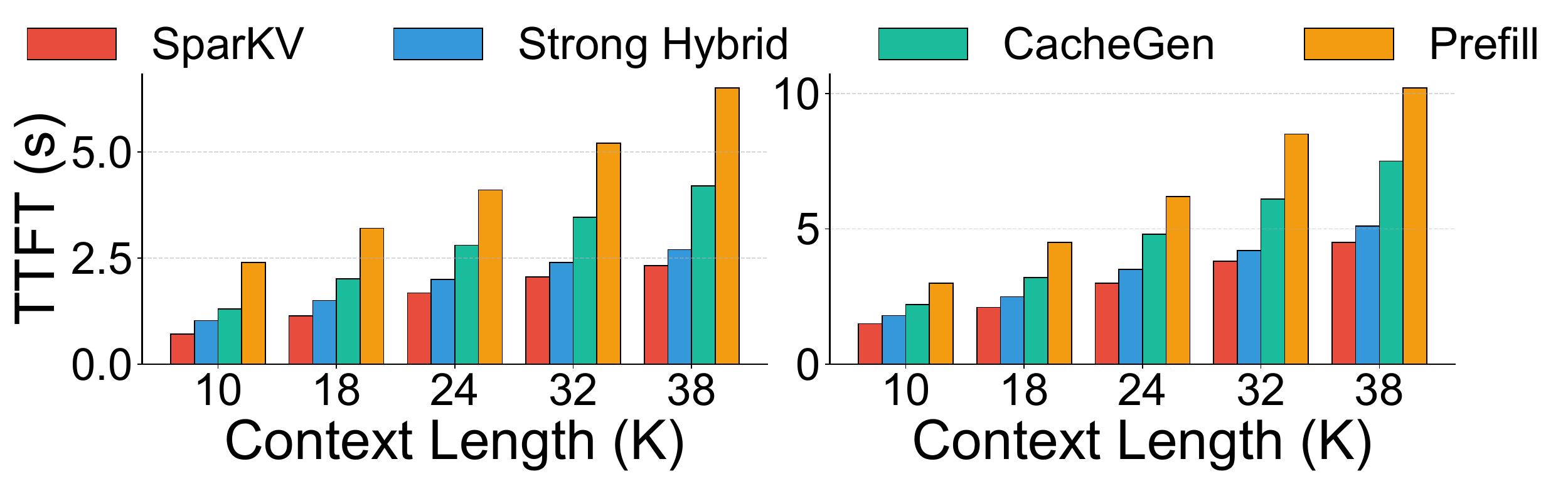}
        \vspace{-0.2in}
        \caption{\small Impact of reusable-context length.}
        \label{fig:context_ttft}
        \vspace{-0.2in}
    \end{minipage}
\end{figure*}

\noindent
\textbf{Robustness to wireless interference.}
To evaluate robustness under volatile network conditions, we introduce controlled access-point congestion using competing devices during LongChat and TriviaQA evaluations. This interference reduces median throughput from 850 to 660\,Mbps under five competing devices, while increasing the standard deviation from 0.25\,Gbps to 0.47\,Gbps. As shown in Fig.~\ref{fig:inter_ttft}, SparKV sustains the lowest TTFT across all interference levels, achieving \(1.4\times\) and \(1.6\times\) speedups over \textit{Strong Hybrid} and \textit{CacheGen}, respectively, under severe congestion. This robustness comes from the runtime adaptation mechanism in \cref{subsec:adapt}: when bandwidth fluctuates, CacheGen's throughput-based bitrate selection becomes less effective, while the static partitioning of \textit{Strong Hybrid} leads to stalls. In contrast, SparKV monitors runtime network conditions and dynamically shifts delayed chunks from streaming to local computation.

\noindent
\textbf{Performance under concurrent requests.}
We further evaluate SparKV on LongChat when edge resources are shared by multiple simultaneous LLM-agent requests. As shown in Fig.~\ref{fig:req_ttft}, SparKV remains stable under high concurrency, with TTFT increasing by only 0.15\,s. In contrast, the baselines degrade much more sharply. Under the heaviest load, SparKV achieves TTFTs that are \(1.4\times\) and \(22.6\times\) lower than those of \textit{Strong Hybrid} and \textit{Local Prefill}, respectively. In addition, SparKV keeps the end-to-end energy per request below 173\,J, corresponding to \(1.5\times\) and \(3.3\times\) reductions relative to the same baselines. Since all methods use the same decoding setup, these energy gains primarily reflect the lower overhead of context preparation and the reduced resource contention enabled by adaptive scheduling.

\noindent
\textbf{Scalability with reusable-context length.}
We evaluate SparKV on reusable contexts ranging from 10K to 38K tokens on the laptop and Jetson AGX platforms. As shown in Fig.~\ref{fig:context_ttft}, SparKV exhibits near-linear TTFT scaling, whereas \textit{Local Prefill} and \textit{Strong Hybrid} suffer from super-linear growth due to the increasing overhead of local attention computation. CacheGen remains bounded by wireless throughput, especially under fluctuating network conditions. In contrast, SparKV dynamically shifts work between local computation and streaming, allowing it to better sustain low TTFT as the reusable context grows.

\begin{figure}[t]
    \centering
    \includegraphics[width=0.9\linewidth]{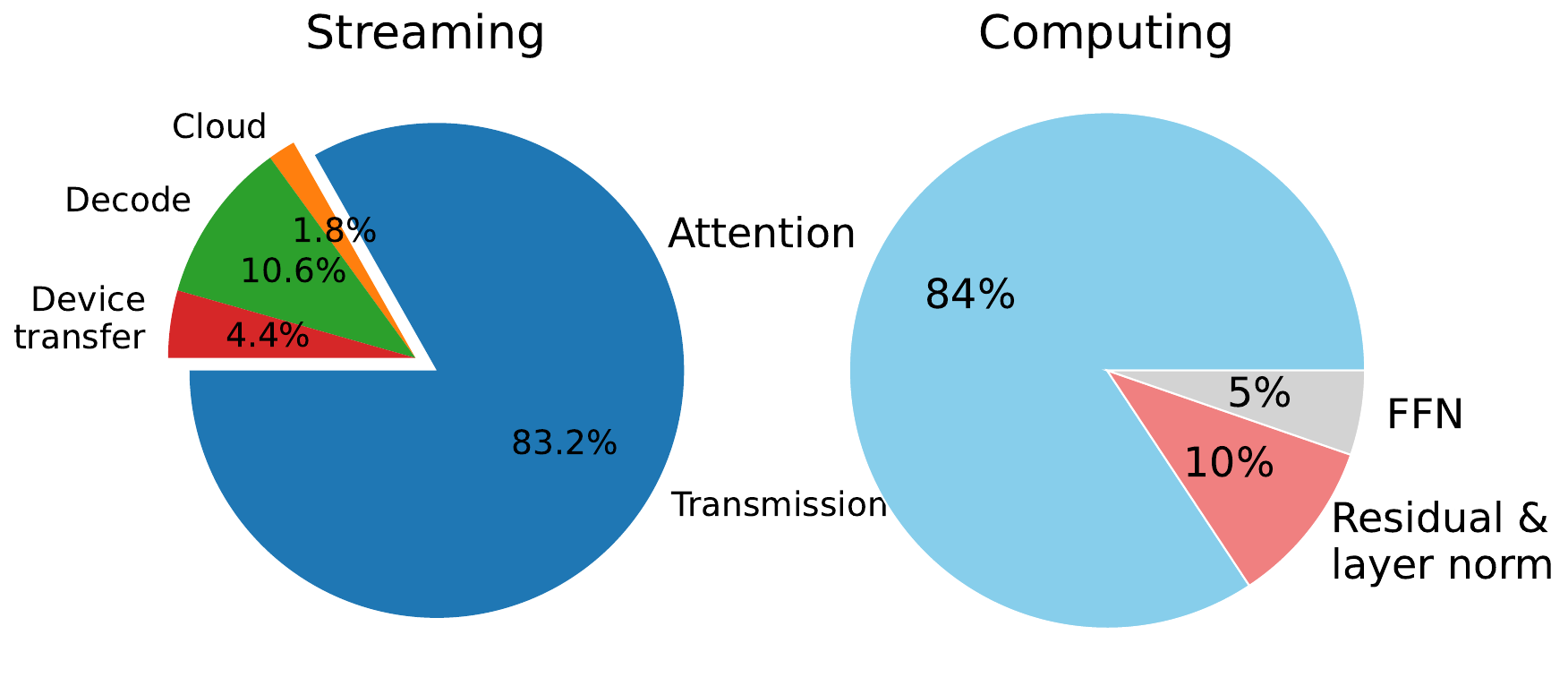}
    \vspace{-0.2in}
    \caption{\small Breakdown of streaming and computation overhead in SparKV on TriviaQA using an RTX 5080 laptop GPU.}
    \label{fig:breakdown}
    \vspace{-0.2in}
\end{figure}

\noindent
\textbf{Overhead breakdown.}
Although our energy metric covers the full request lifetime, most variation across methods comes from context preparation near the start of decoding. We therefore break down KV streaming and local computation overheads and study runtime adaptation under real Wi-Fi traces. Fig.~\ref{fig:breakdown} shows that, for cloud-to-edge KV loading, transmission dominates streaming overhead (85\%), while Huffman decoding and device transfer contribute 14\%. On the computation side, attention accounts for 84\% of local prefill overhead, confirming block-sparse attention as the main optimization target. The remaining operators scale more regularly with sequence length and are modeled as a small scheduler offset.

\section{Related Work}

\noindent
\textbf{On-device LLM deployment.}
Deploying LLMs on edge devices is limited by memory, compute, and energy constraints. Prior work addresses these challenges through model compression and system optimization, such as AWQ~\cite{icml24awq} for weight quantization and recent systems~\cite{asplos25fast} for heterogeneous execution across CPUs, GPUs, and NPUs. \textsc{SparKV} is complementary: instead of compressing weights or redesigning execution, it improves context preparation by jointly scheduling wireless KV streaming and on-device computation.

\noindent
\textbf{KV cache loading and reuse.}
A growing body of work reduces redundant prefill through KV cache reuse and remote loading. InfiniGen~\cite{lee2024infinigen} speculatively prefetches important KV vectors based on query semantics, mainly for local cache reuse. CacheGen~\cite{liu2024cachegen} enables remote KV delivery through bitrate-adaptive KV encoding, while hybrid schemes~\cite{computeorload} overlap KV streaming with local recomputation. In contrast, \textsc{SparKV} explicitly models chunk-level heterogeneity in both streaming and computation and performs dependency-aware scheduling to better exploit overlap under wireless and edge-resource variability.

\noindent
\textbf{KV cache compression.}
Another line of work reduces KV cache size to lower storage and transmission overhead. H2O~\cite{zhang2023h2o} retains heavy-hitter tokens with high prompt attention scores, and LLMLingua~\cite{jiang2023llmlingua} compresses long contexts through selective pruning. Other methods use quantization or low-rank decomposition~\cite{chang2025xkv}; for example, LoRC~\cite{zhang2024lorc} applies progressive layer-wise low-rank compression, and KIVI~\cite{liu2024kivi} uses asymmetric quantization for keys and values. \textsc{SparKV} is orthogonal to these approaches and can naturally incorporate KV compression to further reduce wireless transmission and storage cost.

\section{Discussion and Limitations}

\noindent
\textbf{Support for Multiple Contexts.}
\textsc{SparKV} currently focuses on accelerating context preparation for a single reusable context. However, emerging workloads such as multi-document retrieval-augmented generation often require multiple contexts to be loaded and processed jointly. Treating each context independently can introduce redundant data transfer, repeated KV loading, and fragmented execution. A promising extension is to support multi-context KV reuse and blending, similar in spirit to CacheBlend~\cite{yao2025cacheblend}, so that shared KV states can be reused or merged across related contexts. We leave this direction to future work.

\noindent
\textbf{Extension to Mobile NPUs.}
\textsc{SparKV} is guided by general scheduling principles and is not inherently tied to GPU execution. However, the current implementation and most evaluations are built on GPU-oriented software stacks and sparse-attention kernels. Although our measurement study includes a mobile NPU platform to illustrate the efficiency tradeoffs of KV streaming and local computation, we have not yet fully implemented and evaluated \textsc{SparKV} on mobile NPUs. Extending the system to NPU-oriented runtimes and kernel libraries is an important direction for future work.

\section{Conclusion}
In this paper, we present \textsc{SparKV}, a cloud-edge collaborative framework for KV cache preparation in on-device LLM inference with reusable contexts. \textsc{SparKV} combines wireless KV streaming with local computation through overhead-aware, dependency-aware chunk scheduling, and further adapts the schedule online to handle fluctuations in network and edge resource availability. Extensive experiments across diverse datasets, models, and edge platforms show that \textsc{SparKV} significantly reduces TTFT and energy per request while preserving response quality. These results demonstrate that \textsc{SparKV} is a practical and effective solution for accelerating context-reuse inference on resource-constrained edge devices.

\section*{Acknowledgments}

This work is supported in part by the Research Grants Council (RGC) of Hong Kong under General Research Fund under Grant 11220225. The authors declare no conflicts of interest. 

\bibliographystyle{IEEEtran}
\bibliography{Reference}


\end{document}